\documentclass{article}[12pt, a4]
\usepackage{epsf}

\begin{document}

\title{Fluctuation of the ambipolar equilibrium in magnetic perturbations}

\author{F. Spineanu and M. Vlad \\Association Euratom-NASTI Romania, \\National Institute for Laser, Plasma and Radiation Physics, \\P.O.Box MG-36, Magurele, Bucharest, Romania}
\maketitle

\begin{abstract}
We draw attention on the fast oscillatory deviations from the residual
non-ambipolarity in the case where electrons are driven by strong magnetic
perturbations.
\end{abstract}

If the turbulent diffusion of the electrons and ions are different one can
expect that a radial electric field develops (such that the ambipolarity is
reinstored) and that poloidal plasma spin-up can occur. However it has been
shown that the diffusion of particles in turbulent fields is ambipolar. In
Ref.\cite{Itoh} the difference in the radial fluxes generated by fluctuating
fields has been considered and the r\^{o}le of the ion polarization drift
has been underlined. The radial ion flux partially cancels the radial
electron flux and the ambipolarity is reestablished approximately, to the
order $1/\varepsilon _{\perp }$, which is a very small quantity ($%
\varepsilon _{\perp }$ is the plasma transversal permitivity constant). The
problem has been examined in detail for instabilities having a substantial
magnetic component in Ref.\cite{Waltz}. When the unstable modes that drive
the radial transport are bounded inside plasma (\emph{i.e.} it can be
assumed that there is no exchange of momentum with the region outside
plasma) the fluxes are ambipolar in average over a spatial region including
several mode-resonance surfaces. Pointwise non-zero radial currents, even
reduced by the fraction $1/\varepsilon _{\perp }$, are able to build up
charge filaments and can cause the plasma to be Kelvin Helmholtz unstable.

We return to this problem to examine a situation where the non-ambipolarity
is switched on by an external source acting on the electron component. We
assume that the electrons are radially driven by a magnetic perturbation
which imposes a constant radial flux starting at $t=0$. The ion polarization
drift re-establishes the equality of the fluxes (to order $1/\varepsilon
_{\perp }$) but in this particular case a stationary equilibrium cannot be
reached. The existence of a radial non-zero current yields a non-zero time
derivative of the radial electric field, \emph{i.e.} a non-zero time
derivative of the poloidal velocity. The plasma would be accelerated without
limit in the poloidal direction, but the torque competes with the magnetic
pumping, a very effective damping mechanism. We draw attention to the
oscillatory variation of the velocity, which occurs an a scale much faster
than the dissipative decay.

Such ``events'', consisting of sudden creation of non-ambipolar fluxes,
followed by a fast plasma response in view of reinstoring the ambipolarity,
can appear in a random space and time sequence and on the average can affect
the plasma dynamics. To look in detail to only one event we shall adopt an ``%
\emph{initial value}'' point of view. We consider the slab geometry with $x$
the radial coordinate increasing from the reference point toward the centre
of plasma, $y$ the poloidal coordinate, $z$ is directed along the shearless
magnetic field such as the mixed product of the three versors is positive.
The radial electric current is switched on at $t=0$ , $j_{x}\left(
t=0\right) =-env_{x}^{e}$ (we note $e\equiv \left| e\right| $) and for
simplicity we shall assume that $v_{x}^{e}$ is constant in time and uniform
in space during this single event. From the momentum conservation we find
(with $c$ and $v_{A}$ respectively the light speed and the Alfven speed), 
\begin{equation}
-env_{x}^{e}\approx B_{z}\varepsilon _{0}\left( 1+\frac{c^{2}}{v_{A}^{2}}%
\right) \left( \frac{\partial v_{y}^{i}}{\partial t}\right)  \label{conc1}
\end{equation}
We have neglected the damping and the diamagnetic flow. From the initial
non-ambipolar current $(-env_{x}^{e}=j_{x}\left( 0\right) \neq 0)$ it
results a time-growing poloidal velocity whose magnitude is inverse
proportional to the very large factor representing the plasma transversal
permitivity $\varepsilon _{\perp }=1+c^{2}/v_{A}^{2}$.

Since the time-derivative of the poloidal velocity is constant, the plasma
rotates indefinitely in the poloidal direction with higher and higher
velocity. Naturally this requires the consideration of saturation mechanisms
and makes desirable a time-depending investigation. We write in more detail
the ion momentum equations on $x$ and $y$ 
\begin{equation}
v_{y}^{i}=-\frac{E_{x}}{B_{z}}+\frac{m_{i}}{eB_{z}}\left( \frac{\partial
v_{x}^{i}}{\partial t}\right)  \label{vyex}
\end{equation}
\begin{equation}
v_{x}^{i}=-\frac{1}{\Omega _{i}}\left( \frac{\partial v_{y}^{i}}{\partial t}%
\right) \left[ 1+\frac{1}{\Omega _{i}}\left( \frac{\partial v_{y}^{i}}{%
\partial x}\right) \right] ^{-1}  \label{vx1}
\end{equation}
\begin{equation}
E_{x}=-B_{z}v_{y}^{i}+\frac{m_{i}}{e}\left( \frac{\partial v_{x}^{i}}{%
\partial t}\right)  \label{exvy}
\end{equation}
The charge conservation gives 
\begin{equation}
0=j_{x}+\varepsilon _{0}\frac{\partial E_{x}}{\partial t}  \label{jxex1}
\end{equation}
where 
\begin{equation}
j_{x}\left( t\right) =-env_{x}^{e}+env_{x}^{i}  \label{jx1}
\end{equation}
From these relations and using Eqs.(\ref{vyex}-\ref{exvy}) and (\ref{jxex1}-%
\ref{jx1}) we obtain a single equation for the poloidal ion velocity. 
\begin{eqnarray*}
-env_{x}^{e}\Theta \left( t\right) &=&\frac{en}{\Omega _{i}}\left( \frac{%
\partial v_{y}^{i}}{\partial t}\right) \frac{1}{U}+\varepsilon
_{0}B_{z}\left( \frac{\partial v_{y}^{i}}{\partial t}\right) \\
&&+\frac{\varepsilon _{0}m_{i}}{e\Omega _{i}}\left( \frac{\partial
^{3}v_{y}^{i}}{\partial t^{3}}\right) \frac{1}{U}-\frac{2\varepsilon
_{0}m_{i}}{e\Omega _{i}^{2}}\left( \frac{\partial ^{2}v_{y}^{i}}{\partial
t^{2}}\right) \left( \frac{\partial ^{2}v_{y}^{i}}{\partial x\partial t}%
\right) \frac{1}{U^{2}} \\
&&-\frac{2\varepsilon _{0}m_{i}}{e\Omega _{i}^{3}}\left( \frac{\partial
v_{y}^{i}}{\partial t}\right) \left( \frac{\partial ^{2}v_{y}^{i}}{\partial
x\partial t}\right) ^{2}\frac{1}{U^{3}}-\frac{\varepsilon _{0}m_{i}}{e\Omega
_{i}^{2}}\left( \frac{\partial v_{y}^{i}}{\partial t}\right) \left( \frac{%
\partial ^{3}v_{y}^{i}}{\partial x\partial t^{2}}\right) \frac{1}{U^{2}}
\end{eqnarray*}
where $\varepsilon _{0}$ is the vacuum permitivity, $\Theta \left( t\right) $
is the Heaviside function and 
\[
U\equiv 1+\frac{1}{\Omega _{i}}\left( \frac{\partial v_{y}^{i}}{\partial x}%
\right) 
\]

This equation contains higher time and space $\left( x\right) $ derivatives.
In order to simplify the problem we assume that the externally imposed
electron flux is uniform on the radial direction (for the region of
interest) and this renders the entire model invariant to translation in $x$
coordinate. Then we shall restrict this discussion to solutions which are
independent of $x$. We shall neglect the problem of the stability of the
uniform solutions at perturbations in $x$. Then we have 
\[
-v_{x}^{e}\Theta \left( t\right) =\left( b_{1}+b_{2}\right) w+b_{3}\stackrel{%
\cdot \cdot }{w} 
\]
Here 
\[
w\left( t\right) \equiv \frac{\partial v_{y}^{i}}{\partial t} 
\]
The quantities appearing in the equation have been non-dimensionalized: $%
t\rightarrow t^{\prime }=t/\tau $, $y\rightarrow y^{\prime }=y/l$, $%
v\rightarrow v^{\prime }=v/v_{0}$, where we take as units: $\tau =\Omega
_{i}^{-1}$, $l=\rho _{s}$ the ion Larmor radius at the electron temperature, 
$v_{0}=\beta v_{th}^{e}$ where $\beta $ is the radial magnetic perturbation $%
\widetilde{B}_{x}$ normalised to the main magnetic field, $\beta =\widetilde{%
B}_{x}/B$. $v_{th}^{e}$ is the electron thermal velocity. Finally we remove
the \emph{primes}. The dots means derivative to the time variable. Then the
coefficients are 
\[
b_{1}=\frac{1}{\Omega _{i}}\frac{1}{\tau } 
\]
\[
b_{2}=\frac{\varepsilon _{0}B_{z}}{en}\frac{1}{\tau } 
\]
\[
b_{3}=\frac{\varepsilon _{0}m_{i}}{e^{2}n\Omega _{i}}\frac{1}{\tau ^{3}} 
\]

The solution is obtained by Laplace transform and it reads 
\begin{eqnarray*}
w\left( t\right) &=&\frac{-v_{x}^{e}}{b_{1}+b_{2}}+\left( w_{0}-\frac{%
-v_{x}^{e}}{b_{1}+b_{2}}\right) \cos \left[ \left( \frac{b_{1}+b_{2}}{b_{3}}%
\right) ^{1/2}t\right] \\
&&+w_{1}\left( \frac{b_{3}}{b_{1}+b_{2}}\right) ^{1/2}\sin \left[ \left( 
\frac{b_{1}+b_{2}}{b_{3}}\right) ^{1/2}t\right]
\end{eqnarray*}
The initial conditions are given for the first and the second derivatives of
the velocity, respectively $w_{0}$, and $w_{1}$. In physical units 
\[
w_{0}=-v_{x}^{e}\frac{en}{\varepsilon _{0}B_{z}} 
\]
\[
w_{1}=0 
\]
The frequency of the oscillations is in physical units 
\begin{eqnarray*}
\nu _{osc} &=&\left( \frac{b_{1}+b_{2}}{b_{3}}\right) ^{1/2}=\left( \frac{%
e^{2}n}{\varepsilon _{0}m_{i}}+\Omega _{i}^{2}\right) ^{1/2} \\
&=&\Omega _{i}\left( 1+\frac{nm_{i}}{\varepsilon _{0}B_{z}^{2}}\right) ^{1/2}
\\
&=&\Omega _{i}\varepsilon _{\perp }^{1/2}
\end{eqnarray*}

Now we include the effect of the damping due to the magnetic pumping, $%
\left( \frac{\partial v_{y}^{i}}{\partial t}\right) _{MP}=-\nu v_{y}^{i}$,
where $\nu $ is the appropriate decay rate \cite{Hassam3}, \cite{Haines}, 
\cite{ShaingHirsch1}. The equation restricted to the time domain is now 
\[
-v_{x}^{e}\Theta \left( t\right) =d_{0}v_{y}^{i}+d_{1}\left( \frac{\partial
v_{y}^{i}}{\partial t}\right) +d_{2}\left( \frac{\partial ^{2}v_{y}^{i}}{%
\partial t^{2}}\right) +d_{3}\left( \frac{\partial ^{3}v_{y}^{i}}{\partial
t^{3}}\right) 
\]
where $\Theta \left( t\right) $ is the Heaviside function and 
\[
d_{0}=\frac{\nu }{enB_{z}},\;\;\;d_{1}=\frac{1}{\Omega _{i}}+\frac{%
\varepsilon _{0}B_{z}}{en} 
\]
\[
d_{2}=\frac{\varepsilon _{0}\nu }{e^{2}n^{2}\Omega _{i}},\;\;\;d_{3}=\frac{%
\varepsilon _{0}m_{i}}{e^{2}n\Omega _{i}} 
\]
The solution of the equation is obtained by the Laplace transform and is
written 
\begin{eqnarray*}
v\left( t\right) &=&v_{0}\Theta \left( t\right) +v_{1}q\frac{d_{3}}{d_{0}}+%
\frac{v_{1}c}{\gamma }\exp \left( \gamma t\right) + \\
&&+2v_{1}\left( -\frac{d_{3}\gamma }{d_{0}}\right) \exp \left( \alpha
_{r}t\right) \left[ \left( a_{r}\alpha _{r}+a_{i}\alpha _{i}\right) \cos
\left( \alpha _{i}t\right) -\left( a_{i}\alpha _{r}-a_{r}\alpha _{i}\right)
\sin \left( \alpha _{i}t\right) \right]
\end{eqnarray*}
Here $\alpha $, $\beta =\alpha ^{\ast }$ and $\gamma $ are respectively the
two complex conjugate and the real roots of the polynomial equation $%
s^{3}+\left( d_{2}/d_{3}\right) s^{2}+\left( d_{1}/d_{3}\right) s+\left(
d_{0}/d_{3}\right) =0$. We have for the real $\left( r\right) $ and
imaginary $\left( i\right) $ parts: 
\[
\alpha _{r}=-\frac{d_{2}}{2d_{3}}-\frac{\gamma }{2} 
\]
\[
\alpha _{i}=\left[ -\frac{d_{0}}{d_{3}\gamma }-\left( \frac{d_{2}}{2d_{3}}+%
\frac{\gamma }{2}\right) ^{2}\right] ^{1/2} 
\]
The notations are 
\[
c=\frac{\gamma \beta }{\left( \alpha -\beta \right) \left( \beta -\gamma
\right) }\left[ -1-\frac{\alpha }{\beta }-\frac{q\alpha d_{3}}{d_{0}}-\frac{p%
}{\beta }-\frac{d_{2}}{\beta d_{3}}\right] 
\]
\[
b=\frac{\beta \gamma }{\left( \alpha -\beta \right) \left( \gamma -\beta
\right) }\left[ -1-\frac{\alpha }{\gamma }-\frac{p}{\gamma }-\frac{q\alpha
d_{3}}{d_{0}}-\frac{d_{2}}{\gamma d_{3}}\right] 
\]
and $a=1-b-c$. 
\[
p=\frac{d_{2}v_{1}+d_{3}v_{2}}{d_{3}v_{1}} 
\]
\[
q=\frac{-v_{x}^{e}-d_{0}v_{0}}{d_{3}v_{1}} 
\]

The initial order $i$-derivatives of the velocity are noted $v_{i}$, $%
i=0,1,2 $. They are taken to reflect the sudden onset of the electron flux, 
\emph{i.e.} the initial poloidal velocity is zero, $v_{0}\equiv
v_{y}^{i}\left( t=0\right) =0$ and the first derivative is obtained from the
assumption that at $t=0$ the ions did not yet moved radially ($v_{x}^{i}=0$)
which means that 
\[
v_{1}\equiv \left. \left( \frac{\partial v_{y}^{i}}{\partial t}\right)
\right| _{t=0}=\left( -v_{x}^{e}\right) \left( \frac{1}{\Omega _{i}}+\frac{%
\varepsilon _{0}B_{z}}{en}\right) ^{-1} 
\]
The second derivative is taken zero, $v_{2}=0$. The asymptotic value of the
poloidal velocity is insensitive to this parameter.

\section{Discussion}

\begin{figure}[tbp]
\centerline{\epsfxsize=7cm\epsfbox{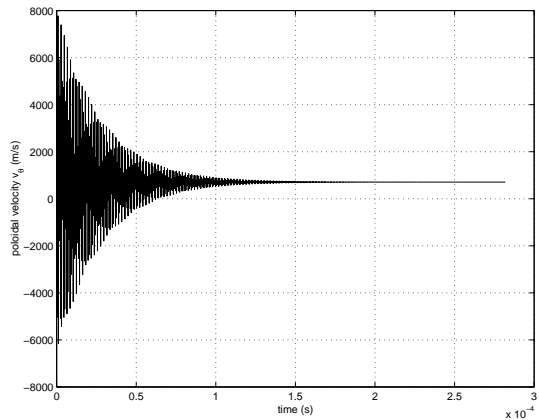}}
\caption{Time variation of the poloidal velocity}
\label{figguerre}
\end{figure}

The most interesting aspect of this solution is the oscillatory behaviour
which is strongly controlled by the non-dissipative part of the equation (%
\emph{i.e.} terms not depending on $\nu $). The dynamic is very clear: just
at $t=0$ the ions acquire a high \emph{time derivative} of their poloidal
velocity, imposed by $j_{x}\left( t=0\right) =-env_{x}^{e}$. The velocity $%
v_{y}^{i}$ starts from zero and grows very fast to values which can be
higher than those required to cancel the non-ambipolar current $j_{x}$. Then
they \emph{reverse} the current and since the electrons are constrained to
follow the external drive (magnetic radial perturbation), the tendency of
rotation reverses and so on. The decay due to the magnetic pumping appears
on longer time scales and this indeed reduces the plasma rotation, without
completely eliminating the oscillations. A plot, Fig. (\ref{figguerre}) of
this evolution is shown, on very large time scale (useful to see the
atenuation).

The asymptotic value of the poloidal velocity is (with our choices of
initial conditions) 
\[
v_{y}^{i}\left( t\gg \gamma ^{-1}\right) =\left( -v_{x}^{e}\right)
enB_{z}\nu ^{-1}
\]
We can see that the nature of these oscillations is similar to that caused
by a deviation form electric neutrality, when the frequency is the \emph{%
plasma frequency} $\omega _{p}=\left( e^{2}n/\varepsilon _{0}m_{e}\right)
^{1/2}$. The important effect of the of these oscillation is in the energy
balance. The plasma rotation takes energy from the source of the initial
magnetic perturbation and from the electron kinetic energy, since both
support the radial electron current. This energy is dissipated via the
collisional ion viscosity represented by the magnetic dumping. A more
complex treatment of this problem should include a detailed electron
dynamics, under the effect of the external magnetic perturbation (which must
be quantified). However, the presence of an oscillatory behaviour is still
expected and suggests to take this into account when the efficiency of
magnetic stochasticity-induced transport is considered.

{\bf Acknowledgments} Useful discussions with J. H. Misguich and R. Balescu are gratefully acnkowledged.

\end{document}